\documentstyle[pra,aps,epsfig]{revtex}
\begin{document} 
 
\title{Experimental Implementation of the Deutsch-Jozsa Algorithm for  
       Three-Qubit Functions 
       using Pure Coherent Molecular Superpositions}

\author{Jiri Vala$^1$\thanks{Present address: Department of Chemistry,
University of California, Berkeley, CA 94720, USA.},
Zohar Amitay,$^2$\thanks{Present address: Department of Chemistry, 
Technion - Israel Institute of Technology, Haifa 32000, Israel.} 
Bo Zhang,$^2$\thanks{Permanent address: Department of Physics,
Royal Institute of Technology, KTH, SE-100 44 Stockholm, Sweden.}
Stephen R. Leone,$^{2}$\thanks{Present address: Department of Chemistry and Department of Physics, and Lawrence Berkeley National Laboratory, 
University of California, Berkeley, CA 94720.}
and Ronnie Kosloff$^1$}
\address{$^1$Fritz Haber Research Center for Molecular Dynamics, 
Hebrew University, Jerusalem, 91904, Israel}
\address{$^2$JILA, National Institute of Standards and Technology  
          and University of Colorado,  
          Department of Chemistry and Biochemistry, and Department of Physics, 
          Boulder, CO 80309-0440, USA} 
  
\maketitle 
 
\begin{abstract} 
The Deutsch-Jozsa algorithm is {\it experimentally} demonstrated for three-qubit 
functions using pure coherent superpositions of Li$_{2}$ rovibrational 
eigenstates. 
The function's character, either constant or balanced, is evaluated by first 
imprinting the function, using a phase-shaped femtosecond pulse, on a coherent 
superposition of the molecular states, and then projecting the superposition 
onto an ionic final state, using a second femtosecond pulse at a specific time 
delay. 
\end{abstract} 


-----------
\vspace{-0.5cm}

\section{Introduction}
\label{sec:introduction}

Quantum computation \cite{Gruska:99,Nielsen:00,Ekert:96}
is aimed at utilizing the quantum nature of physical systems in order to solve computational problems in efficient ways that are impossible in classical computation. 
One of the benchmark quantum algorithms is the Deutsch-Jozsa (DJ) algorithm \cite{deutsch92}.
Its task is to distinguish whether a binary 
$n$-qubit function $f : \{0,1\}^{n} \rightarrow \{0,1\}$
($n$ is given) is constant or balanced.   
The output of a constant function is identical for all possible $n$-qubit inputs, while the output of a balanced function is 1 for half the possible inputs and 0 for the other half. 
The DJ quantum algorithm identifies the constant/balanced character of a function (known to be either constant or balanced) in a single call to the function, 
as compared to the corresponding classical algorithm   
that requires 2$^{n-1}$+1 function evaluations to provide a solution.  
This improvement results from the inherent parallelism when applying a function-dependent unitary transformation on all the possible input elements,  
which are simultaneously contained within a coherent superposition. 
 
In recent years, the Deutsch-Jozsa algorithm, in its revisited form 
\cite{cleve98},  
has been implemented experimentally mainly with NMR techniques. 
Using pseudopure states \cite{gershenfeld97}, it was demonstrated  
for functions having up to a four qubit input \cite{refs_djcleve}. 
%
Also, an implementation of the algorithm for  
two-qubit functions using single photon linear optics has 
been published \cite{takeuchi00}.
A modified version of the algorithm, which does not require a control qubit, 
has been proposed \cite{collins98_version} 
and implemented using NMR for functions with inputs composed of up to 
three qubits \cite{refs_djmodified}.
%
In this version of the algorithm the coherent  
superposition representing the function domain is obtained by applying a first  
Hadamard rotation on the initial state 
$\vert 0\rangle \vert 0\rangle \cdots \vert 0 \rangle$ 
of the $n$-qubit system.
The unitary transformation constructed for each of the various functions introduces a function-dependent phase to each of the elements of the superposition. 
After a second $n$-qubit Hadamard transform is applied, the superposition 
either ends in the initial state if the function is constant 
or in any other state if it is balanced. 
In the present work, this modified algorithm is implemented for  
three-qubit functions using {\it pure} coherent superpositions  
of molecular eigenstates (wave packets). 

Our main motivation behind the current work is exploring the use of a new experimental platform for small-scale 
problem-specific implementation of quantum algorithms. 
The method is based on an ensemble of small molecules in the 
gas phase interacting with a sequence of multiple shaped femtosecond laser pulses. 
The computational task is carried 
out through the time-dependent dynamics of the molecule, with short computation time and very low decoherence rate.
The system is characterized by a value of 10$^{3}$-10$^{4}$ for the ratio between the decoherence time and the 
manipulation time by a shaped femtosecond laser pulse. 
The experimental technique is such that the computation involves {\it pure} states only,
a characteristic that is highly important for a physical implementation of quantum computation. 
These pure molecular states span several (entangled) internal degrees 
of freedom - rotational, vibrational, and electronic. 
 
Currently, we aim at providing a physical platform for a {\it problem-specific} quantum computation. This is different from the common formalism of a 
{\it universal} quantum computation \cite{Nielsen:00}, which is also the one most physical implementations of quantum computation have followed. Under the universal formalism, every unitary transformation acting on $n$ qubits is decomposed into a quantum circuit, which is a sequence of elementary gates, each acting on one or two qubits. The gates composing the circuit are all members within a universal finite set of basic gates.
This model provides a powerful universal programming language for quantum computation. However, it acquires a significant drawback, since most unitary transformations can only be implemented inefficiently, i.e., they require a circuit of elementary universal gates whose size is exponential in $n$. Only special transformations can be decomposed into polynomial-size circuits. One prominent transformation is the quantum Fourier transform, which is also a basis for the most successful quantum algorithm known today, that is the Shor's quantum algorithm for the factorization of numbers \cite{Shor:97}.
Thus, it seems highly desirable that, in parallel to efforts along the above universal quantum computation model, other frameworks for quantum computation will be explored, even if they are non-universal. 
Our objective is to specifically design and shape laser pulses such that under the action of a single shaped pulse the molecules will undergo a complete specific unitary transformation.
This might allow, at a price of losing the universality, the implementation of specific quantum algorithms in a much more efficient way as compared to using other physical platforms designed for universal quantum computation. 
The potential for achieving a complete transformation using a single pulse (or a very small number of pulses) lies in the richness and complexity of the interaction of molecules with a strong broad-bandwidth laser field. 
The corresponding dynamics involves simultaneously single- and multi-photon, direct and Raman, resonant and off-resonant transitions among a large manifold of quantum molecular states.
It is expected that useful theoretical computational methods for designing such pulses will include also optimal control theory \cite{Rice:00}. Initial studies in this direction point to its feasibility \cite{tesch01:cpl,Palao:02}. 
The main limitation of the new platform as implemented in this work is the limited size of the experimentally accessible part of the whole Hilbert space of the internal molecular degrees of freedom. Current experimental techniques provide a basis for handling computational tasks involving pure states with up to hundreds of eigenstates (the equivalent of a Hilbert space of 8-10 qubits) spanning several internal degrees of freedom. This by itself is significant for the experimental research of quantum computation. Moreover, a potential for larger scale quantum computation lies in developing more sophisticated molecular excitation schemes and/or combining the present method with other techniques involving, for example, 
the trapping of cold molecules \cite{Takekoshi:PRL98,DeMille:PRL2002} or cold molecular ions \cite{drewsen:PRA2000}.


\section{Algorithm Implementation}
\label{sec:algo_implement}

\subsection{Using Coherent Molecular Superpositions}
\label{sec:using_coh_super}

In the current implementation of the Deutsch-Jozsa algorithm, 
$2^{n}$ rovibrational eigenstates are being used to represent the eigenstates  
of a system of $n$ entangled qubits. 
Each $n$-qubit function is represented by a different unitary operation stored 
in an oracle, and the task is to determine whether the unitary operation applied 
to the molecular system corresponds to a constant or balanced function. 
The various unitary operations are implemented using phase-shaped broadband 
femtosecond pulses. The interaction of the shaped pulse with a molecule,  
which is initially prepared in a pure single rovibrational state, 
transforms the molecule into a corresponding coherent superposition 
of rovibrational states. Different functions result in different superpositions. 
The physical features 
of the molecule-field interaction allow us to apply the first Hadamard rotation 
and the subsequent unitary transform, representing one of the functions, 
in a single step. The former is associated with the amplitude  
transfer and the latter corresponds to the phase modification: 
$\hat{U}_{phs}\hat{U}_{amp}\vert\psi(t=0)\rangle$. 
The evaluation of the function's character is accomplished by probing  
the overall molecular superposition through its projection onto an ionic 
final state. 
This is accomplished by ionizing the molecule at a single given time delay after  
excitation using a second (unmodified) femtosecond pulse. 
The ability to probe the overall superposition directly as a whole at 
one time,  
which originates from the quantum nature of the process,  
allows the readout step of the algorithm to be achieved without applying  
the second Hadamard rotation. 
The algorithm is performed on an {\it ensemble} of molecules.  
Overall, only one (encoding/imprinting) unitary operation and one measurement  
suffice to find out the function's character. 

\begin{figure}[t]
\centerline{\scalebox{0.35}{\includegraphics{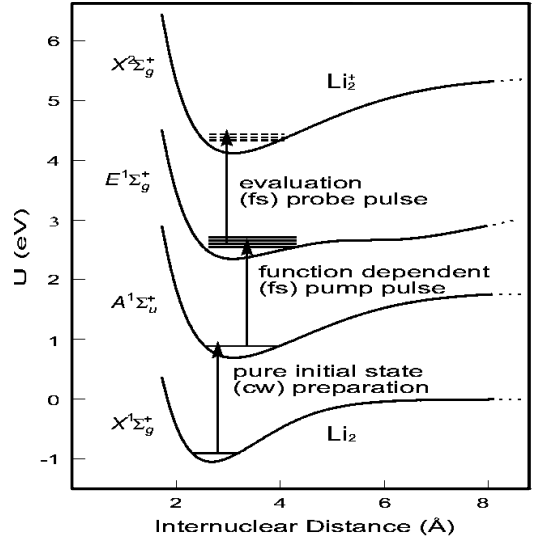}}}
\caption{  
Schematic picture of the excitation scheme of the experiment with  
the relevant potential energy curves of Li$_2$ and Li$_2^+$ [23,24]. 
The rovibrational structure of the E$^{1}\Sigma_{g}^{+}$ state is used for  
the implementation of the Deutsch-Jozsa algorithm.} 
\label{fig:system} 
\end{figure}

 
Although, in general, $2^{n}$ molecular levels are sufficient to represent the Hilbert space of $n$ qubits, in the present experiment $2^{n}+1$ states were used. They include $2^{n}$ rovibrational levels in the electronic E-state of Li$_{2}$ and a single initial rovibrational level in the electronic A-state 
(see Fig.~\ref{fig:system}).
This simplifies the implementation of the unitary operations without any principal limitation. 
Here, the electronic correlation between the  
E-state and the A-state is not probed \cite{pesce01:jcp}, and, thus, the Hilbert  
space naturally reduces to the $2^{n}$ levels in the E-state. 
The algorithm is demonstrated here with three-qubit functions, $n=3$, 
thus eight rovibrational states are employed in the E-state. 
Each rovibrational level, $(v_{E},J_{E})$, corresponds to an eigenstate of the  
three-qubit product space, denoted as 
$|k\rangle \equiv |m\rangle |n\rangle |o\rangle$ with $m, n, o = \{0, 1\}$  
where $k$ is the decadic representation of a three-bit binary digit. 
Explicitly, this means that in the present experiment  
             $|v_{E}=13,J_{E}=17\rangle \equiv |0\rangle \equiv |000\rangle$,  
             $|v_{E}=13,J_{E}=19\rangle \equiv |1\rangle \equiv |001\rangle$, 
             $|v_{E}=14,J_{E}=17\rangle \equiv |2\rangle \equiv |010\rangle$, 
             $|v_{E}=14,J_{E}=19\rangle \equiv |3\rangle \equiv |011\rangle$, 
             $|v_{E}=15,J_{E}=17\rangle \equiv |4\rangle \equiv |100\rangle$, 
             $|v_{E}=15,J_{E}=19\rangle \equiv |5\rangle \equiv |101\rangle$, 
             $|v_{E}=16,J_{E}=17\rangle \equiv |6\rangle \equiv |110\rangle$, 
and  
             $|v_{E}=16,J_{E}=19\rangle \equiv |7\rangle \equiv |111\rangle$.

The unitary transformation, which represents a constant/balanced function, is induced in the molecule by the shaped laser pulse via the Hamiltonian  
\begin{eqnarray} 
\hat{H} =  
\left(\begin{array}{cccc} 
\hat{H}_g & c_{0}^{*}\Omega_{0}^{*} & c_{1}^{*}\Omega_{1}^{*} & \cdots  \\ 
c_{0}\Omega_{0} & \hat{H}_{e0} & 0 &  \\ 
c_{1}\Omega_{1} & 0 & \hat{H}_{e1} &  \\ 
\vdots &  &  & \ddots 
\end{array} \right) 
\; . 
\label{eq:ht_eg} 
\end{eqnarray} 
The symbols $\Omega_{k} = \mu_{k} \epsilon_{k}(t) exp(-i\omega_{k}t)$ ($k=0-7$) 
are the amplitudes acquired by the rovibrational levels on the E-state 
following their excitation by a transform limited (i.e., phase-unshaped) 
pump pulse from the initial rovibrational level on the A-state. 
Each amplitude $\Omega_{k}$ is determined by the transition dipole moment 
$\mu_{k}$ 
between the initial state and the excited state $k$, 
and by the spectral magnitude $\epsilon_{k}(t)$ of the field at the specific 
excitation 
frequency $\omega_{k}$ between these two states.
The time-dependence of the spectral magnitude reflects the pulsed character of 
the field.
The quantities $c_{k} = exp(-i \phi_{k})$ include additional phase factors, 
$\phi_{k}$, that are introduced to the excited rovibrational level 
by the phase shaping of the pump pulse. 
In the current description the influence of off-resonant coupling is neglected. 
The $\hat{H}_{ek}$ are the field-free Hamiltonians of the various excited 
rovibrational levels, and they are equal to the energies of the levels,  
i.e., $\hat{H}_{ek} = \hbar \omega_{k}$. 
Similarly, $\hat{H}_{g}$ corresponds to the initial level on the A-state. 
Under weak field conditions, the excited wave function 
on the electronic E potential 
can be formulated by first order perturbation theory 
\cite{pesce01:jcp,uberna99:faraday,amitay01:chem_phys}.
After eliminating the A-state dynamics by putting $H_g = 0$, 
the excited superposition on the E-state is given at time $\tau$  
after excitation, when the pump pulse is over, as \cite{pesce01:jcp} 
\begin{equation} 
| \psi_e (\tau) \rangle \propto \sum_{k=0}^{7} 
e^{-i \phi_k}  \mu_k \epsilon_k e^{ -i \omega_k \tau}  
| k \rangle \; .
\label{eq:psi_e1}
\end{equation} 
As noted above, $k \equiv (v_{E},J_{E})$ 
while $|k\rangle$ represents the rovibrational eigenfunction 
of state $k$ on the E-state. 
The molecule-field interaction excites each molecular level with a phase and 
amplitude  
that are controlled experimentally by shaping the excitation pulse.
 
Following the formulation of the modified Deutsch-Jozsa algorithm 
\cite{collins98_version}, 
the present pulse shaping is carried out to correspond to a function $f$ 
such that 
\begin{equation} 
| \psi_e (\tau) \rangle \propto \sum_{k=0}^{7} 
a_k e^{ -i \omega_k \tau}  
(-1)^{f(|k\rangle)} | k \rangle \;, 
\label{eq:psi_e2} 
\end{equation} 
where $a_k$ denotes the complex amplitude for a level $k$ 
on the electronic E-state,   
which is independent of the specific evaluated (constant/balanced) function. 
Note, not all the $a_k$ are necessarily equal. 
The term $(-1)^{f(|k\rangle)}$ introduces the function dependent phase factor. 
The 0 and 1 values of the function 
are encoded as a phase of $0^{\circ}$ or $180^{\circ}$ 
(i.e., a $+1$ or $-1$ factor), respectively. 
As a result the expansion in first order perturbation theory,  
$\hat{U} = \hat{1} - i\hat{H}t$,  
provides a model for arbitrary phase and amplitude transfer, 
$\hat{U} =  \hat{U}_{phs}~\hat{U}_{amp}$, 
from the initial A-state level onto the E-state rovibrational wave packet.  
It can be viewed as follows:
\begin{eqnarray} 
& & \hat{U} = \hat{U}_{phs}~\hat{U}_{amp} = \\ 
& & \left(\begin{array}{cccc}
1 & 0 & 0 &  \cdots  \\ 
0 & (-1)^{f(|0\rangle)} & 0 &  \\ 
0 & 0 & (-1)^{f(|1\rangle)} &  \\ 
\vdots &  &  &   \ddots 
\end{array} \right)
\left(\begin{array}{cccc} 
1 & -a_{0} & -a_{1} & \cdots  \\ 
a_{0} & 1 & 0 &  \\ 
a_{1} & 0 & 1 &  \\ 
\vdots &  &  &  \ddots 
\end{array} \right) . 
\nonumber
\end{eqnarray} 
The right term of the {\it r.h.s.}, $\hat{U}_{amp}$, represents the function 
independent 
part, 
while the left term on the {\it r.h.s.}, $\hat{U}_{phs}$, is the unitary 
operation that  
encodes the function values as phases. 
The use of perturbation theory does not represent any principle limitations, 
since the theory of coherent control, for a closed quantum system  
of a discrete spectrum, ensures that 
complete control can always be achieved even with strong fields 
\cite{full_control_ref}. 
 
Each three-qubit binary function is given as a set of eight binary (0 or 1) 
values, 
each corresponding to a possible state, 
$|k\rangle \equiv |m\rangle |n\rangle |o\rangle$, 
of a three-qubit input. 
There are 72 three-qubit constant/balanced functions - 2 constant and 70 balanced.
As mentioned above, a function's value of 1 is represented by a phase value 
of $180^{\circ}$ and a function's value of 0 by a phase value of 0. 
Following the oracle operation for a given function, a set of eight phases is 
determined. Those phases are then encoded experimentally  
into the pump pulse that excites the molecular superposition.  
In practice, in the current experiment, we have chosen to encode these phase 
values  
into the phase-shaped pump pulse as  
an increment over a basic initial set of phases applied to the 
excited rovibrational states,  
$\Phi^{(0)}=\{\phi_{v_E,J_E}^{(0)} ; v_{E}=13-16 ,J_{E}=17,19\}$. 
This $\Phi^{(0)}$ set of phases is actually part of the  
function independent $a_{k}$ coefficients introduced above. 
As a result of this procedure, the two constant functions, 
$f_{1}=\{0,0,0,0,0,0,0,0\}$ and $f_{2}=\{1,1,1,1,1,1,1,1\}$,  
are represented by molecular superpositions having $\Phi^{(0)}$ and 
$\Phi^{(0)}+180^{\circ}$, 
respectively, as their $\phi_{k}$ set of phases (see above).
Since the measured pump-probe ionization signal is sensitive only  
to the relative phases between the various wave packet components,
i.e., insensitive to a global phase of the wave packet,  
both constant functions correspond to the  same pump-probe signal.  
The specific set of phases $\Phi^{(0)}$ used here is  
$\Phi^{(0)} = \{ 
 \phi_{13,17}^{(0)}=298.1^{\circ},  
 \phi_{13,19}^{(0)}=352.0^{\circ},  
 \phi_{14,17}^{(0)}=215.9^{\circ},  
 \phi_{14,19}^{(0)}=137.9^{\circ}, 
 \phi_{15,17}^{(0)}=169.7^{\circ}, 
 \phi_{15,19}^{(0)}=337.6^{\circ},  
 \phi_{16,17}^{(0)}=192.1^{\circ},  
 \phi_{16,19}^{(0)}=0^{\circ} 
\}$. 
These values were chosen such that the ionization of the corresponding 
Li$_{2}$ wave packet at 5 ps delay time after its excitation will result in a {\it global} maximum of the measured coherent signal. 
This is based on previous detailed coherent studies we have conducted on Li$_{2}$ \cite{uberna99:faraday,amitay01:chem_phys}.
The 5 ps time was chosen arbitrarily.   
The balanced functions ($f_{3}$ to $f_{72}$) will result in rovibrational 
wave packets having  
sets of relative phases that are different from $\Phi^{(0)}$, such that the corresponding amplitudes of the wave packet ionization 
signals at  5 ps delay time will be significantly lower than the global  
maximum signal.
Hence,
the identification of the function's character can be made by measuring the 
signal amplitude at this single predetermined delay time,
after a calibration control experiment to measure the global
maximum signal (i.e., for the wave packet with $\Phi^{(0)}$) is initially 
performed one time only.


\subsection{Experimental Technique}
\label{sec:exp_tech}

The molecular excitation scheme of the experiment,
with the relevant potential energy curves of Li$_{2}$ and Li$_{2}^{+}$,
is shown in Fig.~\ref{fig:system}.
The experiment \cite{uberna99:faraday,amitay01:chem_phys} 
is conducted in a static cell, heated to 1023 K, that contains 
a lithium metal sample with Ar buffer gas at about 3.7 Torr (493.3 Pa). 
Using a cw single-frequency dye laser,     
an individual state-to-state transition, from a rovibrational level in the ground X$^{1}\Sigma_{g}^{+}$ state to a rovibrational level in the excited A$^{1}\Sigma_{u}^{+}$ state, is induced on some of the thermally-populated ground state lithium dimers. 
As a result, a small fraction of the overall molecular ensemble is populating 
the single rovibrational state A$^{1}\Sigma_{u}^{+}(v_{A}=14,J_{A}=18)$ as a pure initial state. 
The information that represents a function is encoded into a 
phase-shaped femtosecond laser pulse using a pulse shaping setup.
It employs a liquid crystal spatial light modulator (128 independent pixels) to control the spectral phase of the various frequency components of the pulse.
The wavelength resolution of the pulse shaping setup is $\sim$4.2 cm$^{-1}$ per pixel, and the accuracy of a phase (0-360$^\circ$) applied with the spatial light modulator is better than 1$^{\circ}$, depending on the particular phase value.
The interaction of this shaped pulse with the Li$_{2}$ molecules,    
populated in the selected initial $(v_{A}=14,J_{A}=18)$ state, 
results in a function-dependent tailored coherent superposition (wave packet)
on the E$^{1}\Sigma_{g}^{+}$ electronic state (pump step), which for  
three-qubit functions is composed of $(v_{E}=13-16 ,J_{E}=17,19)$  
(eight states). 
The wave packet excitation can be described within the weak field limit  
\cite{pesce01:jcp}.

Using a second (unshaped) femtosecond pulse, 
this function-dependent excited wave packet is probed after a pre-selected delay time through the ionization of the molecule.
The resulting pump-probe photoionization signal (ion and electron current)
is the measured experimental quantity. 
It is measured using a pair of biased electrodes located inside the lithium cell and a lock-in detection system. All the laser beams (cw, pump pulse, and probe pulse) are periodically modulated using mechanical choppers. 
The measured signal contains a well-identified component that originates only from those molecules that have undergone excitation and ionization due to the timely ordered absorption of three photons: the first from the cw laser, the second from the pump laser pulse, and the third from the probe laser pulse. 
Overall, although most of the molecular ensemble is thermally populating the ground electronic state of Li$_{2}$, the experimental technique allows an ionization signal to be obtained that originates only from those molecules that performed the quantum computing operation. The sub-ensemble containing these molecules is in a pure coherent quantum state.
In general, the measured ionization signal is composed of a constant signal level and a part that depends on the pump-probe delay time 
\cite{uberna99:faraday,amitay01:chem_phys}.
Currently, the measurement was averaged over 10$^{4}$ pump-probe sequences at a specific pump-probe delay time.
The pump and probe pulses originate from a Ti:sapphire laser system (200 kHz repetition rate) with $\sim$160 fs duration, 
12430~cm$^{-1}$ central spectral frequency,
$\sim$150 cm$^{-1}$ bandwidth,  
parallel polarizations, 
and energies of $\sim$0.5 and $\sim$1.0 $\mu$J per pulse, respectively. 
 
The experimental conditions are such that 
the decoherence of the wave packet occurs on a time scale longer than 5 ns.
It is primarily due to collisions between the Li$_{2}$ molecules and 
Ar and Li atoms, which result in pure dephasing and/or state-changing 
transitions. 
The decoherence time scale is at least three orders of magnitude 
longer than the excitation process that encodes the quantum information. 


\section{Experimental Results}
\label{sec:results}

\begin{figure}[t]
\centerline{\scalebox{0.35}{\includegraphics{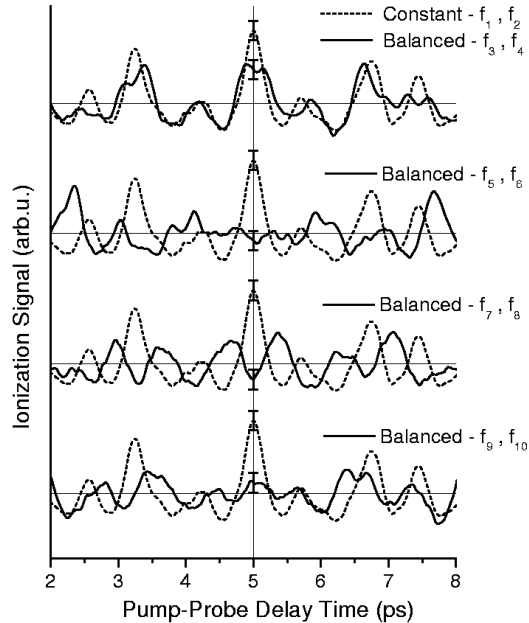}}}
\caption{ 
Pump-probe ionization transients originating from various tailored 
wave packets representing various three-qubit constant or balanced  
functions. The $f_{1}$ and $f_{2}$ are the two constant functions, 
while all the other $f_{i}$ are balanced.
The function's character (constant/balanced) is evaluated by measuring  
the signal level at a single delay time of 5 ps.}
\label{fig:results} 
\end{figure}

Figure \ref{fig:results} displays several pump-probe 
ionization transients out of the complete set of measurements.
They originate from various phase-tailored molecular wave packets, each corresponding to a different three-qubit constant/balanced function. 
For clarity, the transients in the figure are translated vertically to separate one transient from the other. The baselines of all the measured transients, indicated in the figure by the thin horizontal solid lines, are of the same value. 
Similar to the pair of constant functions (see Sec.~\ref{sec:using_coh_super}),
following from the insensitivity of the measured ionization signal to a global phase of the wave packet, 
each transient actually corresponds to either of two functions, 
$f_{i}$ and $f_{j}$, related as $f_{i} = \bar{f_{j}}$, 
i.e., 0 and 1 exchange in their logical representation
(for example, $f_{3}$ and $f_{4}$ given below). 
Such a relationship means that if $f_{i}$ corresponds to a set of phases $\Phi^{i}$, $f_{j}$ corresponds to $\Phi^{j} = \Phi^{i} + 180^{\circ}$, i.e., an additional 180$^{\circ}$ global phase.
The transient shown in all panels by the dashed lines corresponds to the constant functions (f$_{1}$ or f$_{2}$), while the transients shown in thick solid lines each correspond to a different pair of balanced functions. 
The eight balanced functions presented in the figure are
$f_{3} =\{0,0,0,0,1,1,1,1\}$,  
$f_{4} =\{1,1,1,1,0,0,0,0\}$,  
$f_{5} =\{0,1,1,1,1,0,0,0\}$,  
$f_{6} =\{1,0,0,0,0,1,1,1\}$,  
$f_{7} =\{1,1,0,0,0,1,1,0\}$,  
$f_{8} =\{0,0,1,1,1,0,0,1\}$,  
$f_{9} =\{1,1,1,0,1,0,0,0\}$, and  
$f_{10}=\{0,0,0,1,0,1,1,1\}$. 
The complete set of measurements contains results for all the 70 three-qubit balanced functions, measured at 5~ps delay time.
In all cases, as can also be seen in the figure, measuring the coherent signal originating from the ionization of the 
wave packet at the single delay time of 5~ps provides the answer to whether the  
evaluated function is constant or balanced.  
The transients in the figure are shown over the full 2-8~ps delay time range only for clarity.
The signal amplitude at 5 ps that corresponds to 
functions $f_{3}$ and $f_{4}$ is  
the closest one, among all the 70 balanced functions,  
to the signal amplitude at 5 ps of the 
constant functions ($f_{1}$ and $f_{2}$). 
Thus, overall, considering the experimental signal-to-noise levels with the corresponding error bars, the present probability of obtaining a correct answer about the constant/balanced function's character is greater than 99\% for any of the three-qubit functions.
The key ingredient of the read-out procedure is a direct single access 
to the overall set of relative phases encoded in the rovibrational 
superposition. 


\section{Discussion and Conclusions}
\label{sec:disc_conc}

The present implementation of the Deutsch-Jozsa algorithm allows straightforward extension to multi-qubit  
binary functions beyond three qubits. This requires an increase in the  
number of rovibrational levels composing the excited superposition, 
$2^{n}$ levels for $n$ qubits. However, from the way the functions are encoded  
(i.e., always half the levels are phase 0 and the other half 180$^{\circ}$),  
an increase in $n$ does not require an increase in the desired   
experimental signal-to-noise, i.e., there is {\it no} need to ionize more 
Li$_{2}$  
molecules. 
It worth noting that using laser pulses of $\sim$1700 cm$^{-1}$ bandwidth (i.e., $\sim$10-15 fs transform-limited duration) will allow the direct simultaneous coherent excitation, from a single selected Li$_{2}$ rovibrational state of the A-state, of about 50-60 rovibrational states centered spectroscopically around the current excitation region of the E-state.

The present method can be generalized for testing functions that map any even number ($2m$) of inputs into one binary (two-valued) output, and not only $2^{n}$ inputs. We have chosen to work here with functions having $2^{n}$ input values in order to create common grounds with other methods of physical implementation of quantum computation.

In conclusion, we experimentally implemented the modified Deutsch-Jozsa algorithm for three-qubit functions using molecular rovibrational pure states controlled by shaped femtosecond pulses. 
Quantum information representing the function character is encoded into the relative phases of rovibrational levels that constitute the nuclear wave packet on the E$^{1}\Sigma_{g}^{+}$ state of the Li$_2$ molecule. 
The encoded quantum information is measured directly using a weak probe field. This allows, in general, keeping the encoded quantum information primarily undamaged for further quantum operations. This is important in the context of future implementation of more complicated algorithms.

The present task is carried out in the framework of what we consider as coherent parallel computation using quantum elements \cite{coh_comput00}. The quantum nature of the molecule is still not fully utilized for a more efficient computation. 
For example, one quantum property that can be used to advantage is the fact that the molecular eigenstates actually span several internal molecular degrees of freedom - electronic, vibrational and rotational; this might allow conditional manipulation and quantum projection operations.
Nevertheless, this work is a very first step in exploring the {\it experimental} potential of using the present molecular platform for small-scale problem-specific quantum computation. 
The problem-specific approach is different from the universal quantum computation approach. It means that single or very few pulses will be designed specifically to induce complete unitary transformations, without the decomposition into a circuit of basic universal gates that act on one and two qubits. For most unitary transformations, such decomposition is expected to produce a very inefficient quantum circuit. That is, the number of elements in the circuit grows exponentially with $n$.
%
%
By small scale we currently refer to a Hilbert space composed of up to hundreds of quantum molecular eigenstates. This seems to be the experimentally accessible part of the very large complete molecular Hilbert space, which present experimental capabilities can handle.
One route for larger scale computations is the development of much more sophisticated ultrafast molecular excitation schemes, involving, for example, well-controlled Rabi cycling processes. Another route to explore is a hybrid of the present technique, which is based on the manipulation of internal molecular degrees of freedom, with other techniques for the manipulation of external degrees of freedom involving, for example, trapped molecules or molecular ions \cite{Takekoshi:PRL98,DeMille:PRL2002,drewsen:PRA2000}.

This work was supported by the Israel Science Foundation, 
the National Science Foundation, and the Army Research Office.



\newpage 

\end{document}